\documentclass{ws-procs961x669}            
\begin{document}
\title{Challenges in the rheology of glasses}

\author{Peter Sollich$^*$}

\address{Institute for Theoretical Physics,
University of G\"ottingen,\\
Friedrich-Hund-Platz 1, 37077 G\"ottingen, Germany\\
$^*$E-mail: peter.sollich@uni-goettingen.de}

\begin{abstract}
In this contribution to the proceedings of the 29th Solvay Conference on Physics I will give an overview of some key challenges in our theoretical understanding of the rheology of glasses, focussing on (i) steady shear flow curves and their relation to the glass and jamming transitions, (ii) ductile versus brittle yielding in shear startup and (iii) yielding under oscillatory shear. I will also briefly discuss connections to the reversible-irreversible and random organization transitions as well as to the broad field of memory formation in materials.
\end{abstract}


\bodymatter

\section{Introduction and overview}

Rheology is the science of mechanical behaviour and flow. As almost any industrial processing of materials involves deformation, the rheology of supercooled liquids and glasses has evident relevance for applications. What makes the field exciting from a more fundamental point of view is that glasses and other disordered, amorphous materials are already out of equilibrium after preparation by e.g.\ a thermal quench. Deforming them by external forces then generates {\em driven} non-equilibrium systems that have many fascinating features.

In this report I will give an overview of some outstanding challenges in our theoretical understanding of glassy rheology. The field is large, with many connections to other areas, so my coverage will necessarily be selective. The same applies to references to the literature, and I apologize in advance to any colleagues whose work is not cited below. 


 

\section{Steady shear: flow curves and the glass-jamming phase diagram}

\begin{figure}
\includegraphics[width=0.75\textwidth]{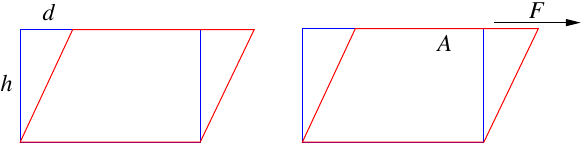} \hfill %
\includegraphics[width=0.18\textwidth]{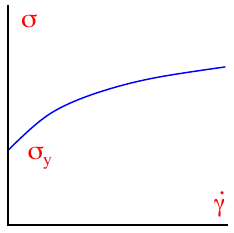}
\caption{Left: Definition of shear strain $\gamma=d/h$. Middle: Definition of shear stress $\sigma=F/A$, where $A$ is the area of the top surface of the block of material. Right: Sketch of a flow curve for a yield stress material, of Herschel-Bulkley form with flow exponent $\beta>1$.}
\label{fig:defs}
\end{figure}

We will focus on shear deformation in the following, which is characterized by the shear strain $\gamma$ and the corresponding shear stress $\sigma$ (see Fig.~\ref{fig:defs}). The task of rheology is then to establish so-called constitutive relations that predict the stress $\sigma(t)$ at some time $t$ after preparation of a material, as a function of the strain history $[\gamma(t')]$ up to that point ($0<t'<t$). This is the situation for strain-controlled experiments; for stress control, the roles of stress and strain in the prediction task are reversed. Either way, deriving such constitutive equations in full generality is a formidable task, so one typically has to focus on specific rheological scenarios.

The simplest such scenario concerns steady shear, where the shear rate $\dot\gamma>0$ is constant. Even if the material initially has a finite shear modulus and resists flow, it will then eventually yield as strains $\gamma$ reach order unity and beyond. In the simplest case a spatially homogeneous flow will result, at a stress that defines the {\em flow curve} $\sigma(\dot\gamma)$. For many glasses, and indeed a wide range of other materials~\cite{Holdsworth93,BonDenBerDivMan17}, this is well described by the Herschel-Bulkley form (sketched in Fig.~\ref{fig:defs} (right))
\begin{equation}
\sigma(\dot\gamma) = \sigma_y + c \dot\gamma^{1/\beta}
\label{HB}
\end{equation}
where $\sigma_y$ is the (dynamic) yield stress below which (even very slow, $\dot\gamma\to 0$) steady flow cannot be achieved. The stress increase at finite $\dot\gamma$ is then governed by the flow exponent $\beta$. At nonzero temperature $T$ one expects the above description to be cut off for slow shear by a Newtonian behaviour $\sigma = \eta_0\dot\gamma$ with viscosity $\eta_0$, but the shear rates $\dot\gamma \lesssim \sigma_y/\eta_0$ where this would be visible are typically too small in the glass phase for this to be measurable (or indeed zero in an ideal glass with $\eta_0=\infty$).

The Herschel-Bulkley flow curve~(\ref{HB}) is always shear thinning at low shear rates, in the sense that the shear rate-dependent viscosity $\eta(\dot\gamma)=\sigma(\dot\gamma)/\dot\gamma$ decreases with $\dot\gamma$. When the flow exponent is in the typical range $\beta>1$, this is true for all $\dot\gamma$. Here and throughout we are presuming that the constituent particles of the glass in question are frictionless; if this is not true, shear thickening rather than shear thinning can occur~\cite{WyaCat14}.

\begin{figure}
\hfill%
\includegraphics[width=0.42\textwidth]{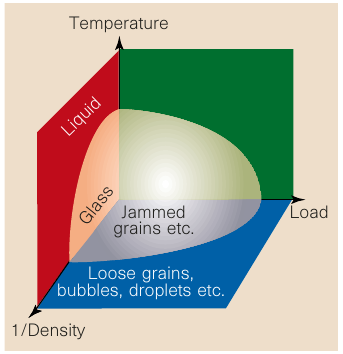} \hfill %
\includegraphics[width=0.45\textwidth]{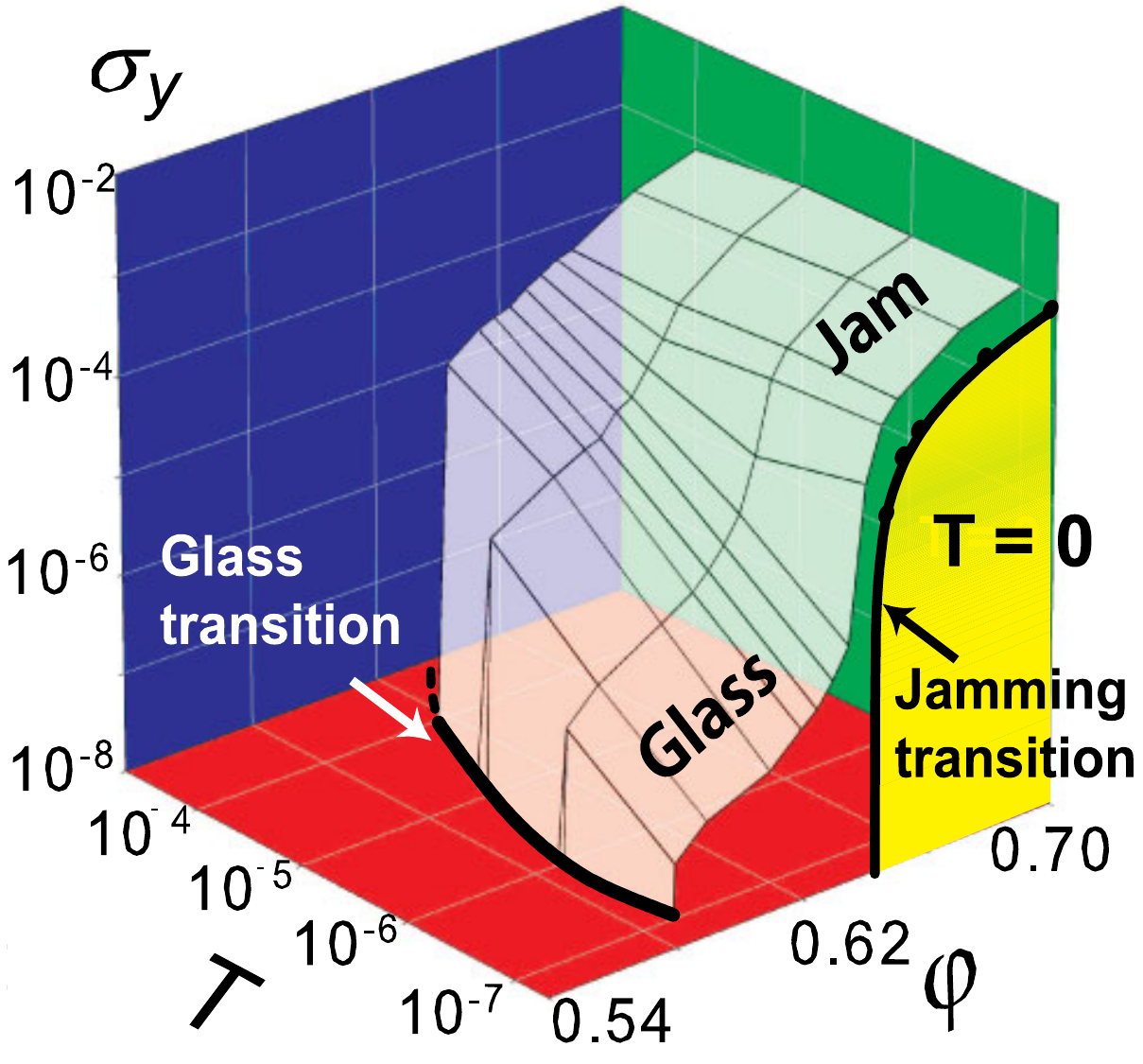}%
\hfill\mbox{}
\caption{Left: Schematic jamming phase diagram from Ref.~\citenum{LiuNag98}, where ``Load'' can be read as stress $\sigma$, with the position of the phase boundary indicating the yield stress $\sigma_y$. Right: Glass-jamming phase diagram from Ref.~\citenum{IkeBerSol12}, from simulations of soft particles with finite-range repulsion. Note the scale-separated low $T$-wing between the volume fractions $\phi_g$ and $\phi_J$ of the glass and jamming transitions.}
\label{fig:jamming_phase_diagram}
\end{figure}

Following Ref.~\citenum{LiuNag98}, one can plot the yield stress against temperature and particle density or equivalently volume fraction $\phi$ to obtain a phase diagram: states at $\sigma<\sigma_y$ are dynamically arrested in the sense that they will not flow (though they may creep, with strain $\gamma(t)$ growing in time but shear rate $\dot\gamma(t)$ going to zero), as shown in Fig.~\ref{fig:jamming_phase_diagram} (left). The seminal idea that flows from this is to view the jamming transition, obtained by compression at $T=0$, and the glass transition upon cooling at constant $\phi$, as part of one unified picture. The jamming transition can be defined as occurring at the volume fraction $\phi_J$ where hard particles cannot be compressed further; however, this does not give a unique value of $\phi_J$ as the result depends on details of the compression protocol~\cite{LiuNag10}. This is consistent with phase diagrams obtained from mean-field theory~\cite{ParZam10,Zamponi_Solvay}, where the jamming transition occurs deep inside the non-equilibrium region. The approach based on the dynamical yield stress $\sigma_y$ as suggested above~\cite{IkeBerSol12,IkeBerSol13} has the advantage that protocol dependences are avoided because one extrapolates from flowing states ($\dot\gamma>0$) that are ergodic -- though with a price to pay in the form of potential uncertainties from estimating the limit $\dot\gamma\to 0$. 

The original schematic jamming phase diagram~\cite{LiuNag98} (Fig.~\ref{fig:jamming_phase_diagram} (left)) suggested that the physics of dynamical arrest is largely governed by the jamming (``J'') point at zero temperature and load. Quantitative numerical studies~\cite{IkeBerSol12,IkeBerSol13} (Fig.~\ref{fig:jamming_phase_diagram} (right)) later showed that the glass and jamming transitions are distinct for particles with a (soft) finite-range repulsion: at low $T$, a glass transition first occurs as $\phi$ is increased to $\phi_g$, with the yield stress jumping discontinuously to a value of order $\sigma_y \sim k_{\rm B}T/a^3$, $a$ being the particle diameter. The scaling with $k_{\rm B}T$ indicates hard sphere-like physics driven by packing entropy, with the particles becoming caged but yet not overlapping. The jamming transition, at a higher volume fraction $\phi_J$, is then signalled by $\sigma_y$ increasing to $O(\epsilon/a^3)$ where $\epsilon$ is the scale of the interparticle repulsion; the rise of $\sigma_y$ as $\phi$ increases beyond $\phi_J$ is continuous on this (at low $T$, much larger) stress scale.


\subsection{Theoretical approaches for the yield stress}

In the following I compare some of the theoretical approaches to the problem of predicting $\sigma_y$ and hence the shape of the glass-jamming phase diagram.

\begin{romanlist}

\item The rheology of soft repulsive particles can in principle be calculated within mean-field theory~\cite{Zamponi_Solvay}, by taking the limit of a large number of spatial dimensions $d\to\infty$. The equations for steady shear have been derived~\cite{AgoMaiZam19a,%
AgoMaiZam19b,Ago21} but have so far resisted numerical solution, so that concrete predictions for flow curves and in particular the yield stress $\sigma_y$ are not yet available (while results for quasistatic shear of finite amplitude do exist, see below).


\item In the framework of mode-coupling theory (MCT), the ``integrating through transients'' (ITT) approach~\cite{FucCat03,BraCatFuc08,
BraVoiFucLarCat09,HajFuc09} predicts that $\sigma_y \approx \sigma_{y,g} + c'(\phi-\phi_g)^{1/2}$ for volume fractions just above the glass transition; the yield stress onset is thus discontinuous, consistent with numerical simulations. The exponent $1/2$ is an effective value obtained from fits to numerical solutions of the MCT-ITT equations, rather than a true asymptotic exponent.

\item Within the random first order transition (RFOT) approach~\cite{Bouchaud_Solvay}, it was argued in Ref.~\citenum{BirBou12} that the yield stress $\sigma_y \sim [G_\infty T\Sigma(T)]^{1/2}$ should be related to the (infinite frequency) elastic shear modulus $G_\infty$ and the configurational entropy $\Sigma(T)$ of the glass. This would imply, rather surprisingly, a vanishing of $\sigma_y$ at the ideal glass transition where $\Sigma(T)=0$.
Also using RFOT arguments, Ref.~\citenum{Lub09} argued instead that only the modulus matters,
 $\sigma_y \sim G_\infty/c''$, with $c''\approx 10^2$ a large geometrical factor. Equivalently $1/c''$ could then be viewed as an estimate for the yield strain $\gamma_y$, around which the stress crosses over from an elastic response $\sigma \approx G_\infty \gamma$ to its steady state value $\sigma_y$ in startup of slow shear. Numerically, however, it is found that the yield stress above the jamming transition has a scaling with $\phi-\phi_J$ close to that of the pressure~\cite{IkeBerSol12,IkeBerSol13}, while $G_\infty$ is larger by a factor of order~\cite{LiuNag10} $(\phi-\phi_J)^{-1/2}$.

\item Elastoplastic models are based on a coarse-grained picture of material elements characterized by a local stress and a yield stress~\cite{NicFerMarBar18}, with possible additional order parameters such as softness~\cite{CubSchRieMalRotDur15,
CubIvaSchStrBasDav17,SchCubKaxLiu17}. Generally such models contain the scale of the yield stress as an input parameter so cannot be used to predict the variation of $\sigma_y$ with $\phi$ and $T$. The same comment applies to shear transformation zone theory~\cite{FalLan98,FalLan11}, which is closely related but distinct in its assumption that local elements have two preferred states (roughly, orientations) and can switch between these when they yield plastically.

\item For mean field versions of elastoplastic models that track distributions of local elastic and plastic properties~\cite{SolLeqHebCat97,Sollich98,HebLeq98,%
AgoBerMarBar15,%
LinWya16}, the onset of a yield stress with the increase of a parameter ($\alpha$ or $x$) characterizing the strength of mechanical noise is typically continuous, e.g.
$\sigma_y \sim (\alpha_c-\alpha)^{1/2}$ for the H\'ebraud-Lequeux (HL) model and $\sigma_y \sim 1-x$ within the soft glassy rheology (SGR) model. 

\item Fluidity models simplify further to characterize the global state of a material in terms of only one or two parameters and have also been used to analyse yield stress properties~\cite{BonDenBerDivMan17}. More sophisticated models with fluctuating local fluidities have also been proposed on the basis of RFOT~\cite{WisWol17}.

\end{romanlist}

\noindent As is clear from the above we do not yet have a theory that can predict yield properties, and in particular the yield stress $\sigma_y$, across both glass and jamming transitions, and this is one outstanding challenge for the future.

\subsection{Flow exponent}

Going to nonzero shear rates $\dot\gamma$, the Herschel-Bulkley form~(\ref{HB}) of the flow curve implies $\dot\gamma \sim (\sigma-\sigma_y)^\beta$, which is analogous to the corresponding relation between velocity and driving force in depinning models~\cite{Wie22}. The power law scaling also suggests that the limit $\dot\gamma\to 0$ is critical, with a plausible underlying physical mechanism being the diverging size of avalanches of plastic activity~\cite{RosSetWya22}. Such critical flow behaviour can in principle occur at generic volume fractions above $\phi_g$ and so is distinct from static jamming criticality~\cite{LiuNag10} at $\phi_J$ itself.

A key factor in the emergence of plastic avalanches is the long-range elastic Eshelby propagator $G(r)\sim r^{-d}$ that describes the perturbation of stresses in the surrounding material resulting from a local plastic relaxation (generally modelled as  incompressible) and the associated stress drop. As the volume integral of $r^{-d}$ is logarithmically divergent, it has been argued that the critical behaviour for $\dot\gamma\to 0$ is of a mean-field nature~\cite{Jag17,FerJag19}. But this conclusion is not trivial on account of the sign-varying quadrupolar structure of the propagator~\cite{EshPei57,PicAjdLeqBoc04}, which contrasts with the positive propagators appearing in depinning~\cite{Jag17}. 

Of the theoretical approaches discussed above, MCT-ITT~\cite{FucCat03,BraCatFuc08,
BraVoiFucLarCat09,HajFuc09} gives an asymptotic flow exponent $\beta=1$ for very small $\dot\gamma$, with larger system-dependent (e.g.\ via details of the interaction potential) effective exponents beyond this. The viscosity estimate from RFOT in Ref.~\citenum{BirBou12} does not yield a flow curve of HB form, but can again be fitted with an effective $\sigma_y$ and $\beta$ across a number of decades. 

Among the mean-field elastoplastic approaches, SGR obtains $\beta=1/(1-x)$ in terms of the mechanical noise temperature $x$ (scaled to be unity at the dynamical arrest transition). The HL model predicts~\cite{HebLeq98} $\beta=2$, while the more recent version~\cite{LinWya18} that accounts properly for the power law decay of the elastic propagator gives $\beta=1$. A scaling picture proposed in the same paper~\cite{LinWya18} gives instead $\beta = 1+1/(d-d_f)$, where $d_f$ is the fractal exponent of the avalanches of plasticity.

The diversity of the above predictions shows a clear challenge in determining whether there are well-defined asymptotic ($\dot\gamma\to 0$) flow exponents at all, what parameters they depend on in case they are not universal, and to what extent they -- rather than only some effective exponent -- can be observed within the range of shear rates accessible to experiment or simulation. More broadly, the universality class of the (potentially) critical dynamical behaviour for $\dot\gamma\to 0$ remains to be determined.

\section{Ductile versus brittle yielding}

\begin{figure}
\hfill%
\includegraphics[width=0.42\textwidth]{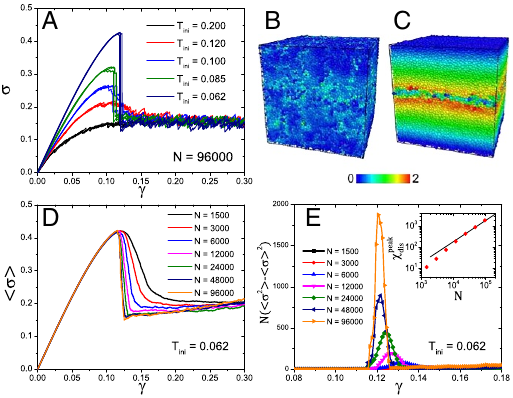} \hfill %
\includegraphics[width=0.47\textwidth]{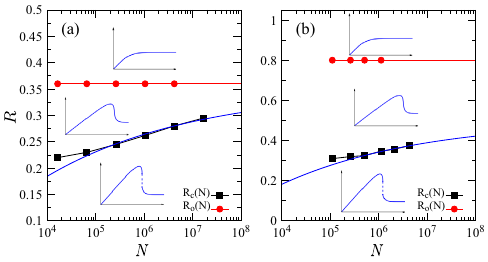}%
\hfill\mbox{}
\caption{Ductile-to-brittle transition in shear startup. Left: Simulation data from Ref.~\citenum{OzaBerBirRosTar18} (A/D: stress-strain curves for different preparation temperatures and system sizes; B/C: snapshots of plastic deformation in ductile and brittle yielding; E: critical behaviour of susceptibility). Right: Dependence of nature of stress-strain curve on system size and initial disorder $R$, from Ref.%
~\citenum{RosBirOzaTarZam22}.}
\label{fig:ductile_brittle}
\end{figure}

We next turn to the behaviour in shear startup, where a nonzero shear rate $\dot\gamma>0$ is switched on at time zero. The steady state that the flow curve describes can then be reached in a {\em ductile} manner, with the flow staying spatially homogeneous and the macroscopic stress $\sigma$ varying continuously in time or equivalently in strain $\gamma=\dot\gamma t$. This variation crosses over from an initially mainly elastic response to the steady flow plateau as the material yields, with a possible overshoot at intermediate strains. If on the other hand yielding occurs in a {\em brittle} fashion, a macroscopic stress drop occurs at some strain, typically accompanied by shear localization~\cite{Fie14,Fie16} into a narrow band, which is akin to a fracture process.

The factors producing ductile versus brittle yielding have been the focus of a significant research effort, two very recent reviews being Refs.~\citenum{DivAgoAimBarBarBen23,BerBirManZam24}. In simulations, a simple and widely used setting is athermal quasistatic shear (AQS), where a system is sheared in small steps followed each time by energy minimization. Here a ductile-brittle transition has been found as a function of the initial disorder in the material~\cite{OzaBerBirRosTar18, PopdeWya18,RosBirOzaTarZam22}, which can be controlled e.g.\ by the annealing protocol used to generate the initial states (Fig.~\ref{fig:ductile_brittle} (left)). Given that the random field Ising model (RFIM) has a similar transition in the variation of magnetization during a magnetic field sweep, a natural conjecture is that the ductile-brittle transition and the associated critical point is in the RFIM universality class. However, again the sign-varying elastic propagator differs from its ferromagnetic counterpart in the RFIM~\cite{RosBirOzaTar23}. This causes important differences between the transitions in the RFIM and in elastoplastic models, already at mean-field level. The statistics of plastic avalanches are a useful probe here that deserves further study; in the HL model, for example, avalanche size distributions indicate that systems under AQS are in a marginal state at all strains~\cite{ParSol23}, which is not the case in the RFIM.

Whether the ductile-to-brittle transition may fall into a new universality class thus remains an open challenge, as does its nature beyond the realm of AQS, where factors such as shear band nucleation timescales and viscoelasticity start to play a role~\cite{DivAgoAimBarBarBen23}. One would also like to understand the location of the transition quantitatively, e.g.\ as a function of initial temperature in a thermal quench. Here the critical temperature has been observed~\cite{OzaBerBirRosTar18} to be close to the $T_c$ of MCT, which may be related to the underlying local yield stress statistics of inherent states~\cite{BarLerLemVanPat20}.

Given that finite size effects are pronounced~\cite{RicRaiLer21} (Fig.~\ref{fig:ductile_brittle} (right)), one should also allow for the possibility that the ductile-to-brittle transition may look rather different for $N\to\infty$. Indeed, arguments based on continuum models~\cite{MooFie13,MooFie14,Fie14,Fie16,BarCocFie20} suggest that finite stress overshoots should generically lead to shear banding instabilities. A smooth stress overshoot, which should occur on the ductile side of an RFIM-like ductile-to-brittle transition (Fig.~\ref{fig:ductile_brittle}), would then be impossible. The resulting transition would therefore need to have a rather different character: the critical case could not have a stress overshoot, and both the stress overshoot and the macroscopic stress drop would have to grow continuously from zero on moving into the brittle regime.

A further challenge in this context is to understand the relation to shear jamming, where for appropriately prepared initial states of hard spheres the stress $\sigma$ diverges at some finite strain during shear startup, a phenomenon that has been seen both in simulations~\cite{JinYos17,JinUrbZamYos18,BabPanJinChaSas21} and mean-field theory~\cite{RaiUrbYosZam15,UrbZam17,AltZam19}.
If slightly soft rather than hard spheres are used so that deformation beyond the shear jamming point is possible, it would be interesting to investigate to what extent the appearance of shear jamming relates to the ductile or brittle nature of the ensuing yielding behaviour. 

\section{Oscillatory shear}

Oscillatory deformations form a further important class of rheological scenarios. We focus here on imposed oscillatory strain (rather than stress), $\gamma(t) = \gamma_0 \cos(\omega t)$. Compared to the constant $\dot\gamma$ cases considered so far, there are now two control parameters, the strain amplitude $\gamma_0$ and the (angular) oscillation frequency $\omega$. For small enough $\gamma_0$ the stress response is then linear and hence also harmonic, and one can think of such rheological experiments as ``mechanical spectroscopy'', probing the linear response of whatever glass has been initially prepared. At larger shear amplitudes $\gamma_0$, higher harmonics with frequencies $3\omega$, $5\omega,\ldots$ come into play. These characterize nonlinear responses~\cite{RouCohHohSolFie08,
RadFie16,
SeyMerCouSieBalWil16,
CarFraFriNicPapVoi17,
LunMicBauLoi17,Fuc17,RadFie18}, with interesting connections to other types of nonlinear spectroscopy, e.g.\ using dielectric measurements~\cite{Bouchaud_Solvay}.

\begin{figure}
\includegraphics[width=0.335\textwidth]{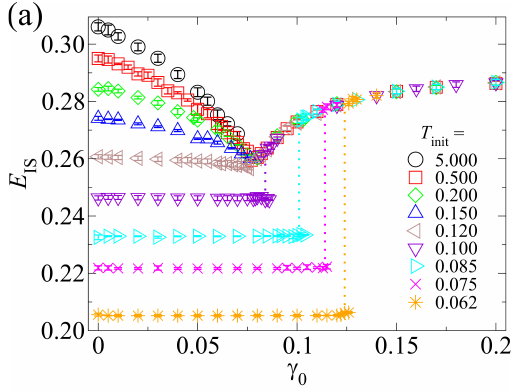} \hfill %
\includegraphics[width=0.29\textwidth]{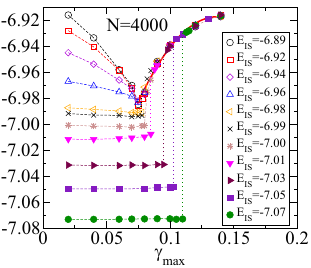} 
\includegraphics[width=0.35\textwidth]{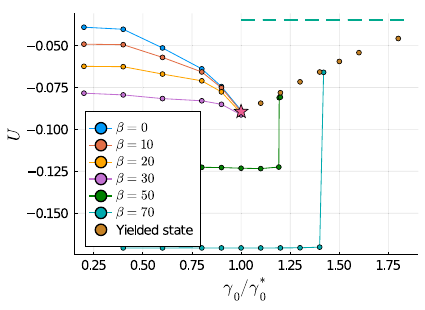}%
\caption{Yielding under slow oscillatory shear strain: Each plot shows steady state potential energy as a function of strain amplitude, with lower curves/symbols relating to better annealed initial states. Left/middle: Simulations from Refs.~\citenum{YehOzaMiyKawBer20,
BhaFofSas21}; right: mean-field model from Ref.~\citenum{ParSasSol22}.}
\label{fig:osc_yielding}
\end{figure}


More generally, shear amplitudes beyond the linear regime are not merely response measurements that explore the state of a glass: the finite driving will modify the properties of the system and may in the long-time limit produce a new periodically driven state -- analogous to Floquet states in quantum mechanics -- that is stationary in a stroboscopic sense. 
There is then again a yielding transition~\cite{
FioFofSas13,
Pri13,
ParSasSol22,
RegLooRei13,
FioFofSas14,RegWebReiDahLoo15,
KawBer16,LeiParSas17,
YehOzaMiyKawBer20,%
BhaFofSas21} as a function of the strain amplitude $\gamma_0$: at small $\gamma_0$ the system settles into an essentially elastic, ``frozen'' steady state while for large strains one finds a liquid-like response. This is somewhat analogous to yielding in shear startup, but conceptually distinct as the transition is one between stationary states rather than in the transient behaviour under shear startup~\cite{ParSasSol22}.

Illustrating the general theme of driven non-equilibrium systems, it turns out that the yielding transition under oscillatory shear has a nontrivial dependence~\cite{%
YehOzaMiyKawBer20,%
BhaFofSas21}
on the initial condition of the arrested states (Fig.~\ref{fig:osc_yielding}): poorly annealed, highly disordered systems reach lower potential energies as $\gamma_0$ is increased so are effectively annealed by the mechanical deformation; beyond the yielding transition at a common strain amplitude $\gamma_0^*$, energies increase again and the fluidization of the system wipes out memory of the initial preparation. Initially well annealed systems, on the other hand, show little if any mechanical annealing and fluidize only at larger strain amplitudes $\gamma_0>\gamma_0^*$. This effect has been reproduced in, amongst others~\cite{SzuGatReg20}, spatially resolved~\cite{KhiTyuVanMal21,%
LiuFerJagMarRosTal22,CocCalFie22} and mean-field~\cite{ParSasSol22} elastoplastic models.

Nonetheless, many open questions remain. These include the nature of the yielding transition, which coming from the fluidized side of large $\gamma_0$ has the character of an absorbing state phase transition but with a potentially exponentially large number of distinct absorbing (elastic, frozen) states; and the role of shear localization, which is seen in simulations at low deformation rates and in continuum models~\cite{RadFie16,RadFie18} but not essential in mean-field elastoplastic models in order to reproduce the observed macroscopic rheological behaviour~\cite{ParSasSol22}. Further challenges arise in understanding the dynamics (with e.g.\ number of shear oscillation periods) and physical manner of the yielding process, which can be non-monotonic in terms of plastic activity~\cite{ParSasSol22} in a manner reminiscent of ``fatigue failure''. There are, in addition, intriguing similarities to step stress experiments that will be worth exploring (see e.g.~Ref.~\citenum{PopDeJiRosWya22}): at constant stress $\sigma>\sigma_y$ the resulting shear rate $\dot\gamma(t)$ can also be non-monotonic in time, dropping to low values before the system eventually yields by starting to flow.

\section{Connections: reversible-irreversible and random organization transitions, memory formation}
\label{sec:connections}

The yielding transition to a fluidized state under oscillatory shear has close connections to a number of other transitions. In particular, it was recognized early on as being linked to the appearance of irreversibility in the underlying particle motion~\cite{FioFofSas13,
Pri13,RegLooRei13,
FioFofSas14,RegWebReiDahLoo15,KawBer16}. This transition in turn has an analogue at lower densities, the random organization transition~\cite{PinGolBraLes05,CorChaGolPin08}. While a glassy, dense material can avoid irreversibility at low $\gamma_0$ by eliminating all ``soft spots'', leading to an elastic response to the imposed strain, in the random organization transition the same is achieved by particles ``isolating themselves'' from each other so that they do not overlap during an oscillatory shear cycle and move affinely with the surrounding solvent. At the transition itself, this leads to hyperuniform structures~\cite{HexChaLev17,%
TjhBer15}.

\begin{figure}
\hfill%
\includegraphics[width=0.6\textwidth]{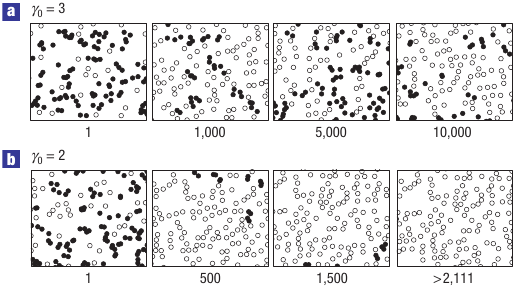} \hfill%
\includegraphics[width=0.341\textwidth]{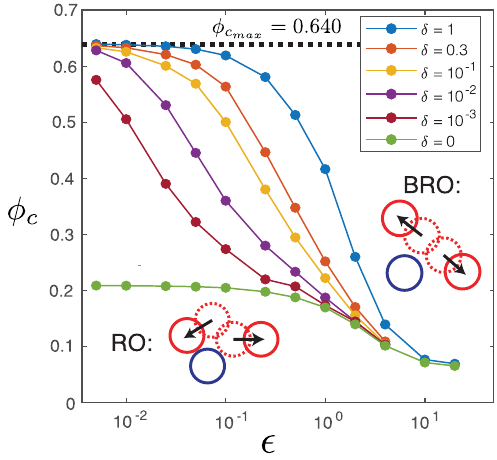} \hfill \mbox{}\\[\baselineskip]
\hfill%
\includegraphics[width=0.385\textwidth]{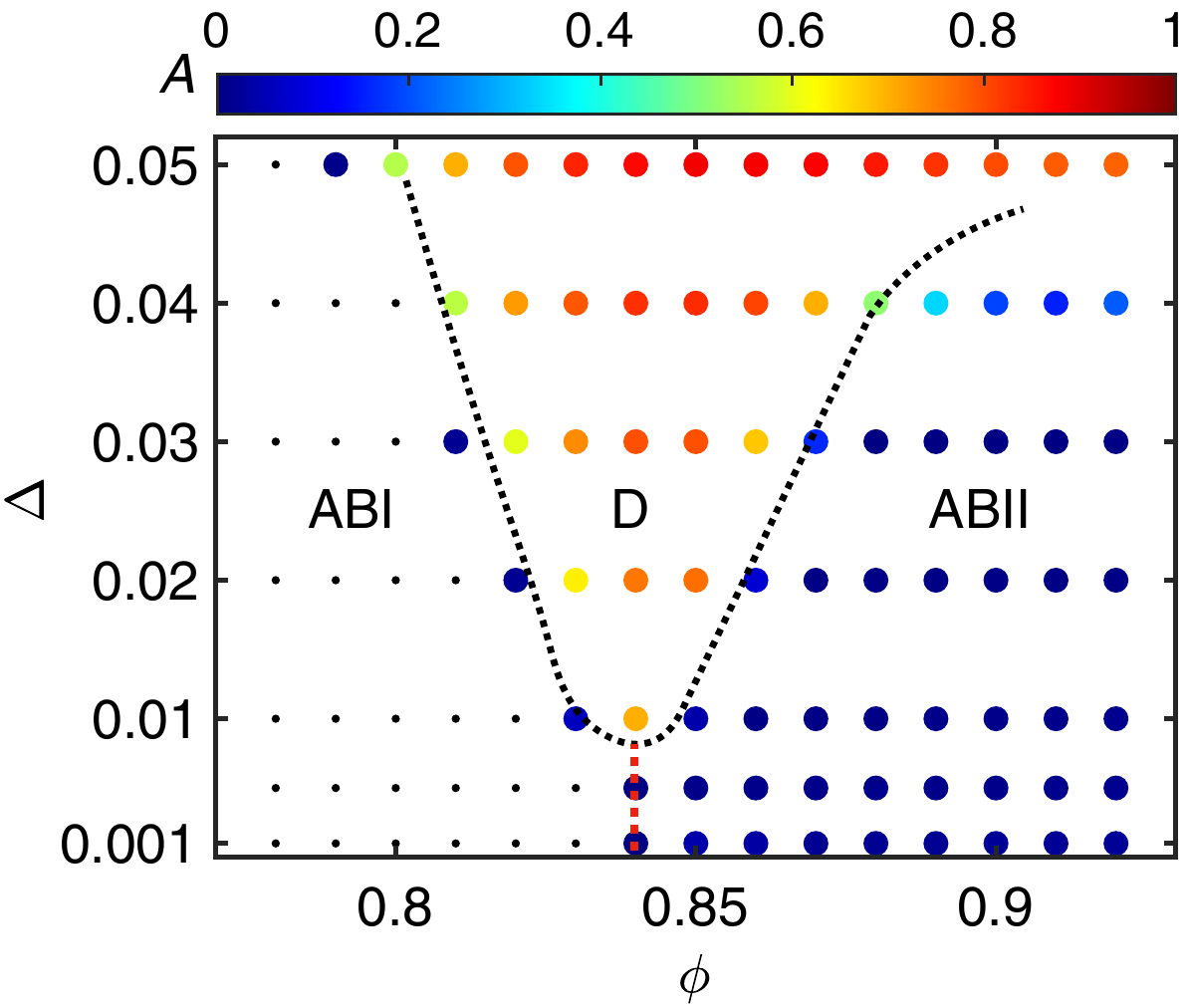} \hfill %
\includegraphics[width=0.47\textwidth]{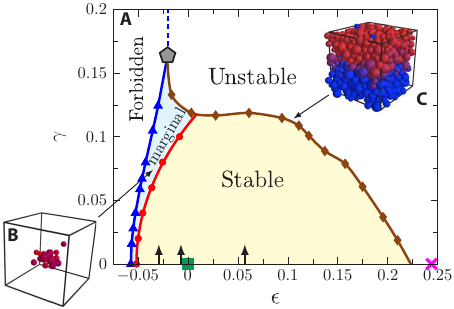}%
\caption{Top left: Illustration of the random organization transition (black particles are active, i.e.\ will collide with another particle in the next strain cycle; number of cycles is shown underneath plots), from Ref.~\citenum{CorChaGolPin08}. Top right: Maximum volume fraction $\phi_c$ reached for absorbing (reversibly evolving) states of the random organization ($\delta=0$) and biased random organization ($\delta >0$) models, with $\epsilon$ setting the scale of random particle displacements in collisions; from Ref.~\citenum{WilGueLevCha21}. Bottom left: Phase diagram of the model of Ref.~\citenum{NesCat20}, where active particles become inactive once they are jammed between too many neighbour contacts; $\Delta$ sets scale of displacements of active particles; reversible motion can arise when particles are isolated (ABI) or jammed (ABII). Bottom right: Compression ($\epsilon$)-shear ($\gamma$) phase from Ref.~\citenum{JinUrbZamYos18}, showing marginal ``partial irreversibility'' region before yield; $\epsilon>0$ corresponds to {\em lower} volume fraction $\phi$.}
\label{fig:rev_irrev}
\end{figure}

Recent studies have shown a highly non-trivial interplay of the various transitions referred to above~\cite{MilSch13,
DasVinSas20,
NesCat20}. It has also been suggested that a biased (towards particles moving away from each other) random organization model~\cite{WilGueLevCha21} creates, as its densest states, random close packed structures with the jamming volume fraction $\phi_J$. This raises in particular the question of the upper critical dimension of the jamming transition: this is generally argued~\cite{LiuNag10} to be $d_u=2$, while the biased random organization model appears to have critical behaviour in the Manna universality class~\cite{WilGuoLevCha23}, with $d_u=4$. A further open question concerns the role of ``partial irreversibility'' before yielding under (quasistatic) oscillatory strain, which has been linked to the marginality of the relevant configurations~\cite{RaiUrbYosZam15,JinUrbZamYos18}.

Viewed from a yet broader perspective, a system under oscillatory strain that does not fluidize can be thought of as memorizing the applied strain amplitude $\gamma_0$. One is thus led to consider more general shear protocols that switch between two or more strain amplitudes, which act as ``training strains'' and can indeed be recalled later by appropriate read-out protocols~\cite{FioFofSas14,AdhSas18,
MukKanSooGan19,
KeiHasKroWie20}. This establishes a fascinating connection between rheological response and the general area of memory formation in materials~\cite{KeiPauZerSasNag19,NagSasZerMut23}, which will definitely be worth exploiting further.

\section{Summary, further challenges and outlook}

In this report I have given an overview of some key challenges I see in our theoretical understanding of glassy rheology. To summarize these briefly, they concern (i) the need for a unified theory for the yield stress covering both the glass and jamming transitions; (ii) an understanding of the critical nature (or otherwise) of steady flow in glasses for $\dot\gamma\to 0$ and of the associated flow exponent; (iii) the universality class and nature of the ductile-to-brittle transition in yielding during shear startup; and (iv) similarly the universality class of yielding in steady state under oscillatory strain and the (even less well understood) associated transient behaviour.

One large area that I have not covered here concerns the interaction of rheology and aging. Linear rheology, for example in step strain experiments measuring a two-time stress relaxation function $G(t,t_{\rm w})$, or conversely in age-dependent creep after small applied step stresses, can act as a probe for the underlying aging dynamics of a glassy system~\cite{Sollich98,FieCatSol09,SolOliBre17,ParFieSol20,ParManSol22}. Nonlinear measurements can give information on the state of aging systems, too, e.g.\ via the age-dependence of the stress overshoot in shear startup or the crossover to steady flow in creep~\cite{FieCatSol09,SieBalVoi12}. In both of these cases, the rheological driving eventually interrupts aging~\cite{Kurchan01}, while conversely oscillatory shear below the threshold for fluidization can potentially drive aging dynamics. 

Significant theoretical progress regarding the interaction of aging and rheological driving has been made in schematic mode-coupling models that can also be motivated from spin-glass physics~\cite{BerBarKur00}, and in mean-field elastoplastic models~\cite{Sollich98,FieCatSol09,SolOliBre17,%
ParFieSol20,ParManSol22}. But there is much scope for more principled theories, with mean-field theory~\cite{Zamponi_Solvay} for $d\to\infty$ an obvious contender. Physically, a clearer understanding of the differences of rheological aging in the glassy and jamming~\cite{SolOliBre17,ChaSolFie19,%
ParFieSol20,ParManSol22} regimes would also be highly desirable. Finally, given the large variety of rheological protocols that could be considered, there is almost certainly further scope in deploying rheology and its interaction with aging as a probe that could help to distinguish between different theories of the glass transition~\cite{Bouchaud_Solvay}.

A more recent addition to the field of glass rheology is the area of dense active matter. Here, like in conventional ``passive'' amorphous materials, dynamical arrest can occur; but breaking of detailed balance at local level by the inclusion of e.g.\ self-propulsion forces brings in a range of new physics. For one, the self-propulsion forces can have their own persistence time $\tau_p$, which raises the question of how the effects of the resulting ``active noise'' compare to those of thermal noise. Similarly, the presence of self-propulsion or other active forces can cause local flows in the material, instead of -- or in addition to -- macroscopic ones generated by external driving. As a result, some very intriguing connections between driving by shear and by activity have been found in recent years~\cite{HenFilMar11,NanGov18,LiaXu18,
MoLiaXu20,ManSol20,ManSol21,ManSol21a,
MorRoyAgoStaCorMan21,VilDur21,
KetManSolJacBer23}, and these will no doubt continue to be the subject of intense study over the next few years. Novel forms of aging have also been discovered~\cite{ManSol20}, with highly persistent active driving effectively allowing relaxation not by activation across barriers, but by fluctuations of the barriers themselves. The consequences of this for the rheological behaviour remain to be explored.

Returning to passive materials, other important issues not covered here include the long-time fate of localized shear ``bands'' arising in yielding, and their relation to ``shear banding'' in the sense of macroscopic regions of a material shearing at different rates, one of which may in fact be zero~\cite{FieCatSol09,Fie14,Fie16}. There are also of course more general deformation modes than the simple shear we have concentrated on throughout, including shear deformations in multiple directions within a single experiment~\cite{KriRamKar23} or isotropic compression and expansion. Compression is natural to consider as it is a standard route to making jammed states, while expansion can lead to cavitation effects; their combination with shear has also been investigated~\cite{
ChaHor16a,JinUrbZamYos18,DatKarCha22}. Finally, while we have focussed on frictionless spherical particles throughout, there is a further wealth of rheological behaviour in glassy systems made up of frictional particles -- to capture important aspects of the physics of granular materials, for example -- or more general particle shapes like hard rods, flexible rods or indeed polymer chains.


\section*{Acknowledgements}

I am grateful to the organizers of the Solvay 2023 conference for the honour of being invited, and the pleasure of taking part in this stimulating and wide-ranging meeting. I am indebted to many colleagues for helpful discussions that have fed into this contribution, in particular Ludovic Berthier and Jean-Philippe Bouchaud as my session chair and co-rapporteur, respectively.


\bibliographystyle{ws-procs961x669}
\bibliography{Solvay_2023_additional,Solvay_2023}

\end{document}